\documentclass[final,3p,times,twocolumn]{elsarticle}

\usepackage{amsmath}
\usepackage{amssymb}
\usepackage{hyperref}

\journal{Physics of the Dark Universe}

\begin{document}

\begin{frontmatter}

\title{Comment on ``Effect of perfect fluid dark matter...'' [Phys. Dark Univ. 34 (2021) 100891]}



\author{Ming-Hsun Li}
\author{Kwei-Chou Yang\corref{cor1}}
\address{Department of Physics and Center for High Energy Physics, Chung Yuan Christian University,
200 Chung Pei Road, Taoyuan 32023, Taiwan}
\cortext[cor1]{Corresponding author. E-mail address: kcyang@cycu.edu.tw}

\begin{abstract}
This Comment clarifies the physical interpretation of the black hole spacetime metric employed by Shaymatov \textit{et al.} [Phys. Dark Univ. 34 (2021) 100891]. We point out that the metric, originally derived in Li and Yang (2012), describes a specific dual-component system consisting of cold dark matter and a phantom field background. Treating this geometry as a single-component ``Perfect Fluid Dark Matter'' (PFDM) with a total equation of state $w_{\mathrm{total}} = -1$ is physically incompatible with the gravitational attraction required for dark matter halos. Furthermore, we clarify the distinct parameter regimes of the original model to prevent potential misapplications in extreme astrophysical scenarios.
\end{abstract}


\end{frontmatter}

\section{Clarification of the Dual-Component Framework}
In Ref.~\cite{Shaymatov:2021}, the authors applied a specific logarithmic metric solution derived in our previous work to perform extended calculations regarding particle motion. However, the physical interpretation of this background requires clarification. 

When summarizing our work, the authors state that we derived a black hole solution \textit{``where the dark matter envelope is described by a bounded phantom scalar field.''} This statement conflates two physically distinct entities. In the original framework (Li and Yang, 2012)~\cite{LiYang:2012}, the galaxy system is explicitly composed of:
\begin{itemize}
    \item \textbf{Cold Dark Matter (WIMPs):} Responsible for gravitational attraction ($w\simeq0$).
    \item \textbf{Phantom Field Background:} Responsible for driving the accelerated expansion of the Universe at cosmological scales (acting as dark energy with $w<-1$). However, at the galactic scale, this background is spatially inhomogeneous and strictly dominated by a negative kinetic energy gradient.
\end{itemize}

Equating a repulsive phantom field with gravitating dark matter obscures the underlying physics upon which the stability proofs were built. The phantom field provides the \textit{background geometry}, but it does not constitute the dark matter particles themselves.

\section{Equation of State and Metric Parameter Constraints}
In Sec.~2 of Ref.~\cite{Shaymatov:2021}, the stress-energy tensor calculated directly from the logarithmic metric (Eq.~(2) therein) yields the relation $p_r=-\rho$ (Eq.~(3) therein). The authors subsequently define this effective fluid as a single component: ``Perfect Fluid Dark Matter.'' This interpretation leads to a conceptual inconsistency. 

First, an effective radial equation of state $w_{r,\mathrm{total}} = p_r/\rho = -1$ indicates a strongly negative radial pressure. Labeling such effective fields simply as ``dark matter'' is fundamentally incompatible with the astrophysical nature of dark matter halos. While typically modeled as non-relativistic matter with negligible macroscopic pressure ($p \simeq 0$), any viable dark matter candidate must provide net gravitational attraction, strictly precluding a strongly negative pressure such as $p_r = -\rho$. This apparent negative pressure is not a property of a single dark matter fluid, but rather an emergent collective behavior of the dual-component system in a highly specific localized regime.

To fully appreciate this mischaracterization, one must understand the exact solutions provided in the original framework. In Li and Yang (2012)~\cite{LiYang:2012}, we derived a class of static, spherically symmetric solutions parameterized by $\epsilon$, defined via the energy-momentum tensor condition ${T^\theta}_\theta = {T^t}_t (1-\epsilon)$. The general solution for the metric function $-g_{tt} = g_{rr}^{-1} \equiv F(r)$ splits into two distinct forms:
\begin{equation}
    F(r) = 1 - \frac{r_s}{r} + \frac{r^{2(1-\epsilon)}}{r_\epsilon}, \quad \text{for } \epsilon \neq \frac{3}{2}, \label{eq:power}
\end{equation}
and a specific mathematical limit,
\begin{equation}
    F(r) = 1 - \frac{r_s}{r} + \frac{a}{r}\ln\left(\frac{r}{|a|}\right), \quad \text{for } \epsilon = \frac{3}{2}. \label{eq:log}
\end{equation}

We demonstrated that the physical regime corresponding to a dark matter-dominated environment (i.e., a galactic halo explaining star motion) strictly requires the power-law metric (Eq.~\ref{eq:power}) with $\epsilon \simeq 1 - \alpha/2 \simeq 1.08$. In this realistic regime, the total energy density is strictly dominated by the cold dark matter field ($\rho_{\mathrm{total}} \simeq \rho_{\mathrm{DM}}$), while the phantom field density becomes negligible ($\rho_{\mathrm{ph}} \simeq 0$). This is the proper configuration to describe a fluid that behaves essentially as dark matter.

In contrast, the logarithmic metric adopted by Shaymatov \textit{et al.} (Eq.~(2) in Ref.~\cite{Shaymatov:2021}) corresponds exactly to our Eq.~(\ref{eq:log}) (with the notational changes $r_s \to 2M$ and $a \to \lambda$). This specific metric is a localized geometric solution resulting from the intense coexistence of WIMPs and the phantom background at the limit $\epsilon = 3/2$, yielding a macroscopic $p_r = -\rho$. In this state, the underlying phantom field is forced into an exotic configuration with $\rho_{\mathrm{ph}}=-\frac{\lambda}{16\pi r^3} < 0$, $p_{r,\mathrm{ph}}=-\frac{\lambda}{8\pi r^3}$, and a local equation of state $w_{\mathrm{ph}} = 2$. This strongly violates the Null Energy Condition (NEC), i.e., $\rho_{\mathrm{ph}}+p_{r,\mathrm{ph}}=-\frac{3\lambda}{16\pi r^3}<0$. Extracting this specific $\epsilon = 3/2$ solution—which fundamentally relies on an exotic phantom component—and broadly relabeling the entire geometric background as a generic ``Perfect Fluid Dark Matter'' obscures the underlying dual-component physics.

\section{Physical Limits of the Spacetime Metric}
Because Ref.~\cite{Shaymatov:2021} identifies the effective fluid as a generic, single-component PFDM, the metric is decoupled from its essential physical constraints. It should be emphasized that the metric they employ is not a universal vacuum solution but a specific geometric manifestation of the stable coexistence between WIMPs and a phantom background field. 

Utilizing this spacetime to discuss extreme physical scenarios (such as infinite magnetic fields or arbitrarily high center-of-mass collision energies in their subsequent sections) requires caution. Without accounting for the underlying exotic phantom dynamics and the necessary effective field theory limits, such extrapolations may lead to unphysical regimes. The metric inherently depends on its dual-component origin and cannot be treated as a generic fluid background without potentially violating the foundational physics that ensures its stability.

\section{Conclusion}
In summary, the metric derived in Li and Yang (2012) describes a specific dual-component interacting system and cannot be treated as a simplified PFDM background without altering its core physical meaning. We hope this mathematical and conceptual clarification assists the community in accurately applying this spacetime solution in future astrophysical studies, thereby preventing the propagation of physical misinterpretations.

\section*{Acknowledgments}
This work was partly supported by the National Center for Theoretical Sciences and the National Science and Technology Council of Taiwan under grant number NSTC 114-2112-M-033-008.

\end{document}